\documentstyle[aps,prb,preprint]{revtex}
\begin{document}
\title
{Quantum-Critical Behavior in a Two-Layer Antiferromagnet}
\author{Anders W. Sandvik\cite{abo}}
\address{National High Magnetic Field Laboratory, 1800 E. Paul Dirac Dr.,
Tallahassee, FL 32306}
\author{Andrey V. Chubukov}
\address
{Department of Physics, University of Wisconsin-Madison,
 Madison, WI 53706 and
P.L. Kapitza Institute for Physical Problems, Moscow, Russia}
\author{Subir Sachdev}
\address{Department of Physics, POB 208120, Yale University,
New Haven, CT 06520-8120}
\date{\today}
\maketitle
\begin{abstract}
We analyze quantum Monte Carlo data in the vicinity
of the quantum transition between a Neel state and a quantum paramagnet
in a two-layer, square lattice spin 1/2 Heisenberg antiferromagnet.
The real-space correlation function and the universal amplitude
ratio of the structure factor and the dynamic susceptibility
show clear evidence of quantum critical behavior at low temperatures.
The numerical results  are in good quantitative agreement with
$1/N$ calculations for the  $O(N)$ non-linear sigma model. A discrepancy,
reported earlier, between the critical properties of the
antiferromagnet and the sigma model is resolved.
We also discuss the values of prefactors of the dynamic susceptibility and
the structure factor in a single layer antiferromagnet at low $T$.
\end{abstract}
\narrowtext

The two-layer, square lattice spin 1/2 Heisenberg antiferromagnet
has recently attracted some attention for a number of reasons.
It has been proposed that its properties should help explain neutron
scattering, photoemission and relaxation rate
data in $YBCO$ and $Bi-2212$  cuprate superconductors~\cite{photo,Mil93,lee}.
On the theoretical side, it is among the simplest, two-dimensional
Heisenberg spin systems which display a quantum phase transition between a Neel
state and a  quantum paramagnet~\cite{Mil93}. Quantum Monte Carlo (QMC)
simulations on this model
do not suffer from a sign problem, making it possible to obtain precision data
on the critical properties of the quantum transition. It therefore constitutes
an attractive setting for testing our understanding of transitions in
two-dimensional Heisenberg spin systems.  This is particularly important
as there are conflicting reports on the observation of quantum-critical
behavior in a single-layer
antiferromagnet~\cite{CHN,CSY,birg,CSS,Elstner,San95}.

A popular approach for the analysis of the collective properties of
two-dimensional, square lattice antiferromagnets has been to model the
long-wavelength, long-time spin fluctuations in terms of those of the $O(3)$
non-linear sigma model~\cite{CHN,CSY}.
 The main difference between the antiferromagnet and the
sigma model is that the microscopic action for the former contains Berry
phases,
associated with the commutation relations between different
components of the Heisenberg spins. The Berry phases however cancel
between
the two sublattices for smooth spin configurations~\cite{berry}. For the
two-layer
antiferromagnet there should be a further cancellation between the two layers,
which should occur even for singular saddle points
(pairs of ``hedgehogs'' on the two layers) of
the effective action. There are, therefore, strong theoretical reasons for
believing that the quantum transitions in the two-layer antiferromagnet and the
sigma model should be in the same universality class.

In a recent paper~\cite{San94} reporting QMC results on the two-layer
antiferromagnet near the critical point, it was found that while the
measurements were qualitatively consistent with the predictions of the
sigma model, there was a discrepancy in a coefficient in the temperature ($T$)
dependence of the correlation length ($\xi$).
The origin of this discrepancy was not understood, and it was not
clear whether the theoretical picture based on the sigma-model
should be questioned.  On the other hand, high temperature
series studies~\cite{Elstner} found good agreement with the predictions
of the sigma-model at somewhat higher temperatures.

In this
paper, we re-examine this issue by improving the accuracy of the numerical
calculations, and performing a more detailed analysis of the data.
As shown below, we
now find complete consistency between the
measured universal critical properties
of the two-layer antiferromagnet and the sigma model, and understand the
origin of the earlier discrepancy~\cite{San94}.
For comparison, we will also present some additional data on the correlation
function in a single layer antiferromagnet  at low temperatures,
when the system is in the renormalized-classical regime.

We start by presenting some new analytical results. Consider the spin
correlation function in real space, $G(r)$, where $(-1)^r G(r)$ is the Fourier
transform of the static structure factor,
\begin{equation}
S(k) = T \sum_{\omega_m}
\chi (k, \omega_m),
\label{S}
\end{equation}
where $\chi$ is the dynamic susceptibility, and $\omega_m$
is a Matsubara frequency. Here and below we set $k_B = \hbar =1$.
Results for $\chi (k, \omega_m)$ in the quantum-critical region
have been presented in Ref 5---using these results,
we obtain at the critical coupling, but $T$ finite
\begin{equation}
G(r) = \frac{1}{3}~N^2_0~\left(\frac{3 T}{2 \pi\rho_s}\right)^{1 + \eta}~
\int_0^{\infty} x
dx ~\Xi (x)~ J_0 \left(\frac{x r T}{c}\right).
\label{gr}
\end{equation}
Here $x$ is a rescaled momentum,  $J_0$ is a Bessel function,
$c$ is the spin-wave velocity, and $\rho_s$ and $N_0$ are
spin stiffness and sublattice magnetization respectively, measured
just away from the critical point on the ordered side---the ratio
$N^2_0/\rho_s^{1+\eta}$ remains finite upon approaching criticality.
Finally, $\Xi (x)$ is the universal
crossover function, which we have computed
in a $1/N$ expansion for the $O(N)$ sigma-model  for
$0<x<5$ (previously~\cite{CSY}, this function was computed only in the limits
of
small and large $x$). We found that to the accuracy needed here, $\Xi (x)$ to
order $O(1/N)$ can be modeled by the same  functional form as in the
$N = \infty$ result,
\begin{equation}
\Xi (x) = \frac{1 + 2 n_x}{2 \sqrt{x^2 + m^2}},
\label{Xim-f}
\end{equation}
where $n_x = (\exp{\sqrt{x^2 + m^2}} -1)^{-1}$ is a Bose function, but
with $m \approx 1.04$  instead of the $N=\infty$, $m_0 =
2 \log[(\sqrt{5} +1)/2] = 0.9624$.
The difference between the $1/N$ result for $\Xi$,
extended to the physical case
of $N=3$,  and eq. (\ref{Xim-f}) is less than 1\% for $x <2$ and does not
exceed 2\% for larger $x$. This form of $\Xi$ indeed implies that the
argument of the Bessel function can be rewritten as $1.04 x r/\xi$
(where $\xi^{-1} = mT/c$).

The two terms in (\ref{Xim-f}) yield the same exponential decay of the
spin correlations at large distances, but
with different prefactors. The term without a Bose function is a purely quantum
contribution, which yields
 $e^{-r/\xi} /r$, while the integration over $x$ in
the second term in (\ref{Xim-f}) yields a classical
result $e^{r/\xi}/(r/\xi)^{1/2}$.
In the earlier report, the QMC data near the
critical point were fitted using only the first quantum contribution to
$G(r)$ at large
distances~\cite{San94}.
Meanwhile, we have checked that
the dominant contribution to $S(k)$ in (\ref{S}) actually
comes from the few first terms in the frequency sum.
Consequently,  the piece in $G(r)$ associated with the
classical fluctuations is at least as important   as the piece related to
quantum fluctuations. Moreover, as the QMC data were obtained for
finite-size systems, we found it useful to extract the correlation
length from the fit  to the functional form of $G(r)$ at moderate distances
of $5-10$ lattice spacings. At such distances,
 $G(r)$ is not necessarily well approximated by the exponential form, and
one should use the full eq.~(\ref{gr}) for numerical comparisons.

The advantage of introducing the correlation length
into the argument of the Bessel function is that eq. (\ref{gr}) no longer
contains the spin-wave velocity which in a two-layer
system is not known precisely~\cite{dirk}.
Once $\xi$ is obtained from the fit to the QMC data, one can use the
QMC results for the uniform susceptibility ($\chi_u$)
and compute the dimensionless ratio
$Q(T) \equiv \xi^{-1} (T) / (T \chi_u (T) )^{1/2}$. It is predicted~\cite{CSY}
that at the critical point, $Q(T)$ takes a $T$-independent universal
value---using the result for $\chi_u$ we obtain
$Q \approx 1.92m = 2.0$ (at $N=\infty$, $Q=1.64$).
Away from the critical point, we expect that as $T\rightarrow 0$,
$Q$ vanishes as $\sim \exp(-2 \pi \rho_s / T)$ on the N\'{e}el ordered side,
and $Q$ diverges as $\sim \exp(\Delta /2 T)$ on the quantum-disordered side
($\Delta$ is the spin-gap).

For the numerical calculations we have used a variant of the Handscomb
QMC technique~\cite{qmc}. All results for two coupled layers shown here are
for systems with $2\times L\times L$ spins with $L=32$. For a single layer,
we discuss results for systems of size $64\times 64$.

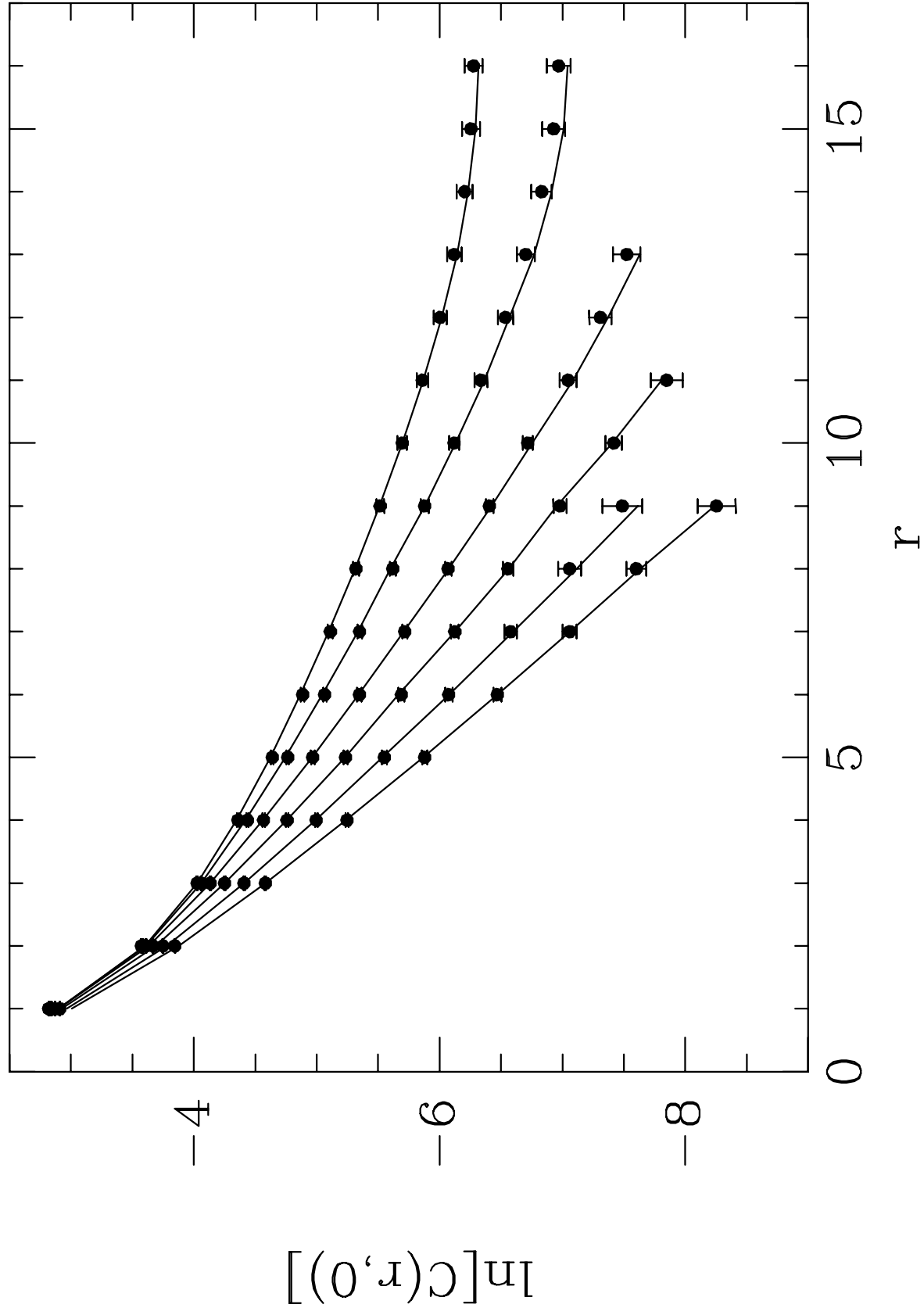
\begin{figure}
\caption{QMC results for the staggered spin correlation function
$G(\vec r)= (-1)^{r}
\langle (S^z_{1, r}-S^z_{2,r})(S^z_{1,0}-S^z_{2,0})
\rangle$ (the indices $1,2$ refer to the two planes) along $\vec r = (r,0)$,
for an interplane coupling $J_2 = 2.55J_1$ at temperatures $T=0.3-0.8J_1$
(the top curve is for
$T = 0.3J_1$). The solid curves are fits to
$G(r)+G(L-r)$, with $G(r)$ given by (\protect\ref{gr}).}
\label{figgr}
\end{figure}

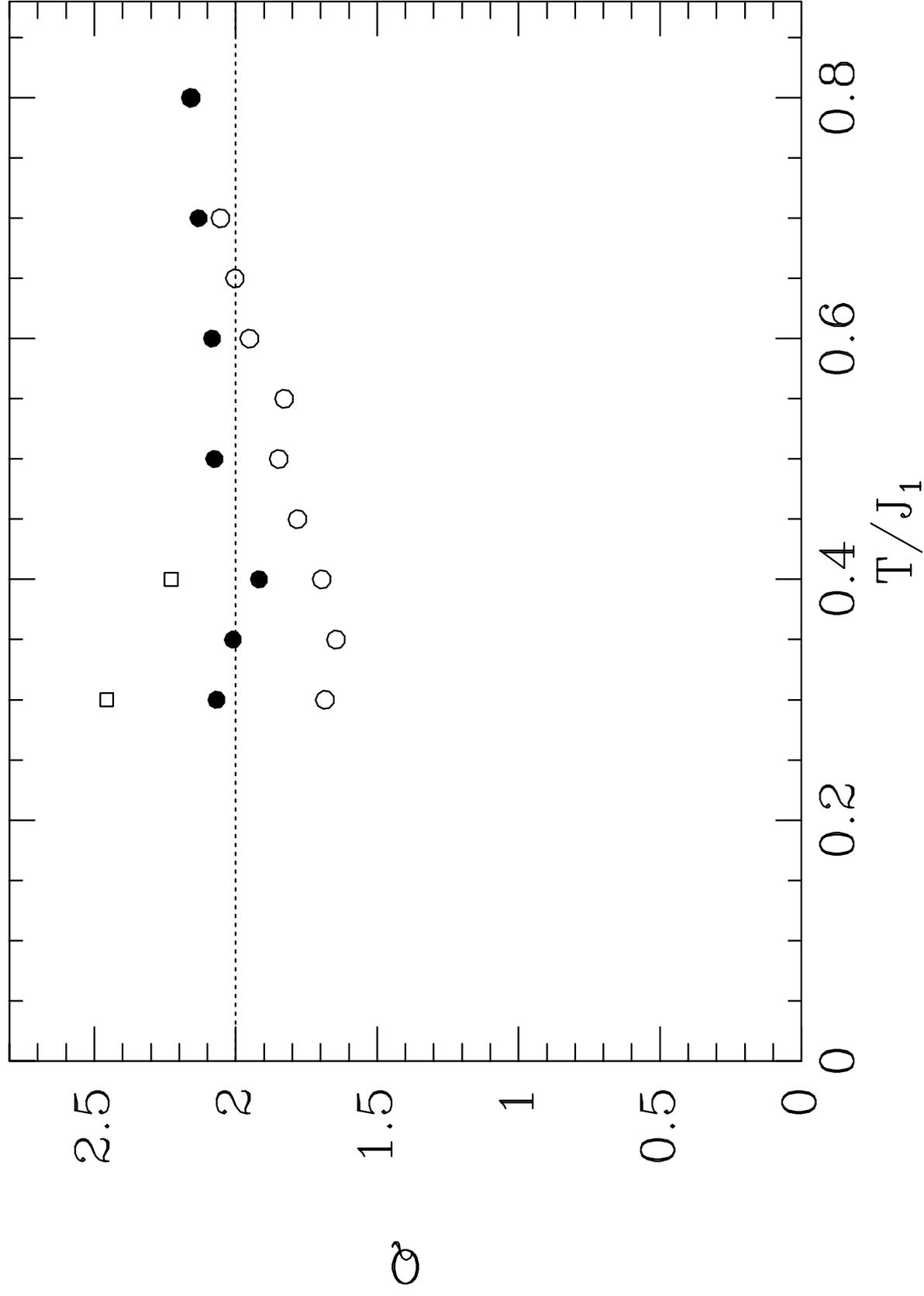
\begin{figure}
\caption{QMC results for the universal parameter ratio
 $Q = \xi^{-1}/(T \chi_u)^{1/2}$ for the interplane
couplings $J_2/J_1=2.50$ (open circles), $2.55$ (solid circles), and $2.60$
(open squares). The dashed line is the prediction for the
quantum-critical regime.}
\label{figQ}
\end{figure}

The  QMC data~\cite{San94,Hid92,San94_2} for $G(r)$ in the two-layer model at
a near critical $J_2/J_1 = 2.55$
are presented in Fig.~\ref{figgr} ($J_2$ and $J_1$
are inter and intralayer exchange integrals, respectively).
We fitted the data for $G(r)$ to eq. (\ref{gr}) with $\xi$ and the prefactor
as the fitting parameters. We then used the QMC results for the uniform
susceptibility~\cite{San94} and obtained $Q(T)$.
The results, along with results for systems
just above ($J_2 / J_1 = 2.6$) and below ($J_2 / J_1 = 2.5$)
the critical point, are shown in Fig~\ref{figQ}.
The results are in good accord with the theoretical predictions discussed
above.
At the critical point $J_2/J_1 = 2.55$, $~Q$ changes little with
temperature below
$T \sim 0.5J_1$ and remains close to the theoretical prediction, $Q = 2.0$,
(i.e, $\xi^{-1} = 1.04 T/c$) {\it{down to the lowest temperature studied}}.
 To the best of our
knowledge, this is the first observation of the $universal$ quantum-critical
behavior of the correlation length in 2D antiferromagnets.

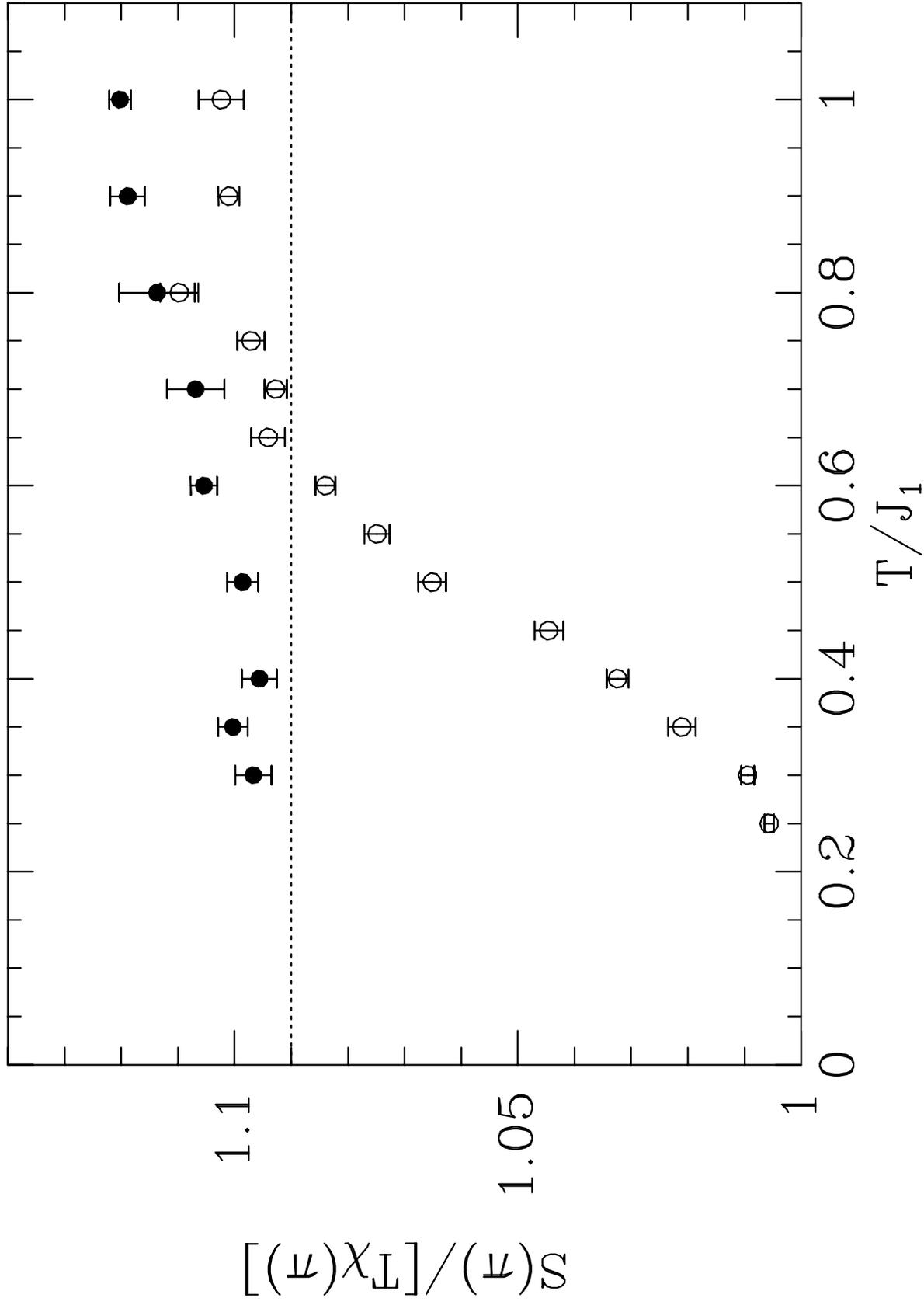
\begin{figure}
\caption{
QMC results for the universal ratio $S(\vec \pi)/[T\chi(\vec \pi)]$  for a
single plane (open circles), and two coupled planes with $J_2/J_1 =
2.55$ (solid circles). The dashed line is the prediction for the
quantum-critical regime.}
\label{figstchi}
\end{figure}

Another interesting dimensionless ratio~\cite{Elstner} of the
quantum-critical region is
 $W(T) \equiv S(\vec{\pi})/T\chi(\vec{\pi},0)$.
On the ordered side, as $T \rightarrow 0$,  $W \rightarrow 1$
as only the first term in the frequency sum in (\ref{S}) is important in this
classical regime. At the
critical point we however have to perform the full frequency summation.
Using the $1/N$ results of Ref 5, we find $W \approx 1.05 m = 1.09$.
The QMC results for the ratio are shown in Fig.~\ref{figstchi}.
 We see that at the transition point, $W$ is  $T$ independent and
it value is close to the predicted one down to the $lowest$ temperatures
studied. Notice that earlier
studies~\cite{Elstner} have found the same value of $W$ in a single-layer
antiferromagnet, but at relatively high temperatures, $T > 0.6J_1$, when
nonuniversal corrections to $\chi$ and $S$ are
relevant, and one can only hope that
 they are canceled out in the ratio $S/T\chi$.

The prefactor
$A = N^2_0 ~(J_1/2\pi \rho_s)~(T/2 \pi \rho_s)^{\eta}$
extracted from the fits to the correlation function
is shown in Fig.~\ref{figa}. We see that $A \approx 0.063$
is almost $T$-independent, as it should be given that $\eta$ is very small.
This prefactor can also be extracted from the product $V (T) \equiv
S(\vec{\pi})\chi_u (T)$ which again does not contain the spin-wave
velocity. In the
$1/N$ expansion~\cite{CSY}, we obtain $V (T) = 1.7 A$. Averaging the QMC
data for $V$ at $T=0.3J_1 -0.5J_1$, we obtain $V \approx 0.107$ which yields
$A = 0.063$ in complete agreement with the above result.

Finally, we extracted $m$ and $A$ directly from the fit of the QMC data to eq.
(\ref{Xim-f}). We used $c = 1.8J_1$ which is the average of
 $c=1.7J_1$ and $c =1.9J_1$ obtained from the fits to the data for the uniform
susceptibility~\cite{San94} and the ratio~\cite{Sand95}
$S(k)/T\chi(k,0)$ at $k \leq \pi$.
Using $m$ and $A$ as fitting parameters, we obtained
$A = 0.063$ and $m = 1.02$;
the latter is again very close to the theoretical prediction.
The $A$ factor was also extracted from
QMC calculations of the static structure factor $S(\vec{\pi})$. At the lowest
available $T \sim 0.3J_1$, we obtained using $c = 1.8J_1$,
 $A = 0.063$, in complete agreement with the above results.

For comparison, we also discuss some new low-temperature
data for a single-layer antiferromagnet, particularly for the
prefactor of the correlation function. At low enough temperatures, the
physics of a single-layer antiferromagnet is governed by thermal, classical
fluctuations. The $N=\infty$ result for $\Xi$ is now $qualitatively$
wrong, as the $1/N$ corrections are logarithmically divergent, and the series
for the $1/N$ terms gives rise to an extra power of $T$ in the structure
factor.
The real space correlator $G(r)$ now has the form
\begin{eqnarray}
G(r) &=& Z_N~\frac{N^2_0 T}{\rho_s}~
\left(\frac{(N-2)T}{2 \pi\rho_s}\right)^{1/(N-2)}\nonumber \\
&&\int_0^{\infty} dx~x
f(x/m)~ J_0 \left(\frac{x r T}{c}\right),
\label{grrc}
\end{eqnarray}
where $f(y)$ is a universal scaling function ($f(0) =1$), and $Z_{N}$ is the
overall renormalization factor. The behavior of $f(y)$ at large $y$ is rather
complex, but  we actually need to know $f(y)$ only for $y=O(1)$ as
the momentum integration in (\ref{grrc})  is confined to $x \sim m$. For such
$y$, virtually all  $1/N$ corrections to the mean-field expression for $f$;
$f(x/m) = (x^2 + m^2)^{-1}$,  can be absorbed into the renormalization of
the mass~\cite{CSY}, so that for practical purposes we again can use the
mean-field functional dependence of $f$. This is consistent with the
earlier analysis of Tyc {\it et al.}~\cite{tyc}, who introduced a
semi-phenomenological one-parameter scaling form  for $f(y)$.
For $y =O(1)$, this scaling form is well approximated by the mean-field
expression. The prefactor $Z_{N}$ was obtained~\cite{cr} analytically
to first order in $1/N$.
 To estimate $Z_3$, we assume that the
$1/N$ corrections collapse into a single exponent (this is known to be
true for the correlation length~\cite{hasen}).
We then obtain
\begin{equation}
Z_N = \left(\frac{\Gamma(1/3) \Gamma (7/6)}{\Gamma(2/3)
\Gamma(5/6)}\right)^{3/(N-2)}~\frac{2^{(2-N)}}{\Gamma
(1 + 1/(N-2))}.
\label{Z}
\end{equation}
To first order in $1/N$, we have~\cite{CSY} $Z_N = 1 +
0.188/N$, while
for $N=3$, we have $Z_3 \approx 2.15$, which is larger than the first-order
result but still substantially smaller than
the previous estimate of Tyc {\it et al.}~\cite{tyc}---$Z_3 \approx 6.63$.

The  QMC data for the real-space correlation function and their fit to
the eq.~(\ref{grrc}) are shown in Fig.~\ref{figrc}. There is no
need to exclude the spin-wave velocity, as in a single-layer antiferromagnet
$c$ and other input parameters are known to high accuracy~\cite{Iga92}.
The mass extracted from the
fit at low temperatures agrees well with the results of other numerical
studies~\cite{DM} and with the theoretical formula for the classical
region~\cite{CHN,hasen}, though perfect agreement with the theory
is reached at slightly smaller $\rho_s$. The prefactor is now
$A = (N^2_0/2\pi \rho_s)~[Z_3~(T/2\pi \rho_s)]$, i.e., it should decrease
linearly with $T$. The $A$ inferred from the fit indeed
decreases with $T$, though there are not enough points at low $T$
to obtain the $T$ dependence. The value
of $A$ at the lowest temperature in the simulations, $T = 0.3J_1$ is $A =
0.07$, while from the theoretical formula
we obtain $A \approx 0.05$. This indicates that the actual value of $Z_3$ is
probably about 40\% higher than in (\ref{Z}). Another way to find $A$ is
to consider the ratio $X = S(\vec{\pi})J^2_1/(T \xi)^2 = N^2_0 Z_3 J^2_1/(2 \pi
\rho_s^2)$. The QMC result for this ratio is $X\approx 1.34$
at $T=0.3J_1$, while the theory gives $X = 0.98$, which is again
less by about 40\%.

To summarize, in this paper we presented the first numerical evidence for the
universal quantum-critical behavior of the correlation length in a 2D
antiferromagnet near the critical point {\it {at low temperatures}}.
We compared the results for the universal parameter ratios
$\xi^{-1}/(T \chi_u)^{1/2}$
and  $S(\vec{\pi})/[T \chi (\vec{\pi},0)]$ with QMC data,
and in both cases obtained  good agreement with the theory based on the
$O(3)$ sigma-model.
The prefactor $A$ extracted from the fit is nearly
$T$ independent which agrees with
the theoretical prediction $A \propto T^{\eta}$.
For comparison, we also presented data for the correlation function of
a single-layer antiferromagnet at low $T$. We found good agreement
with the theoretical result for the correlation length in the
renormalized-classical region, but the prefactor in the correlation
function disagrees with the classical formula by about 40\%.

We thank H. Monien, D. Morr, D. Scalapino and A. Sokol
for useful conversations. A.W.S. was supported by the DOE
under Grant No. DE-FG03-85ER45197 and
the ONR under Grant ONR N00014-93-0495.
S.S. was supported by NSF Grant DMR-9224290.

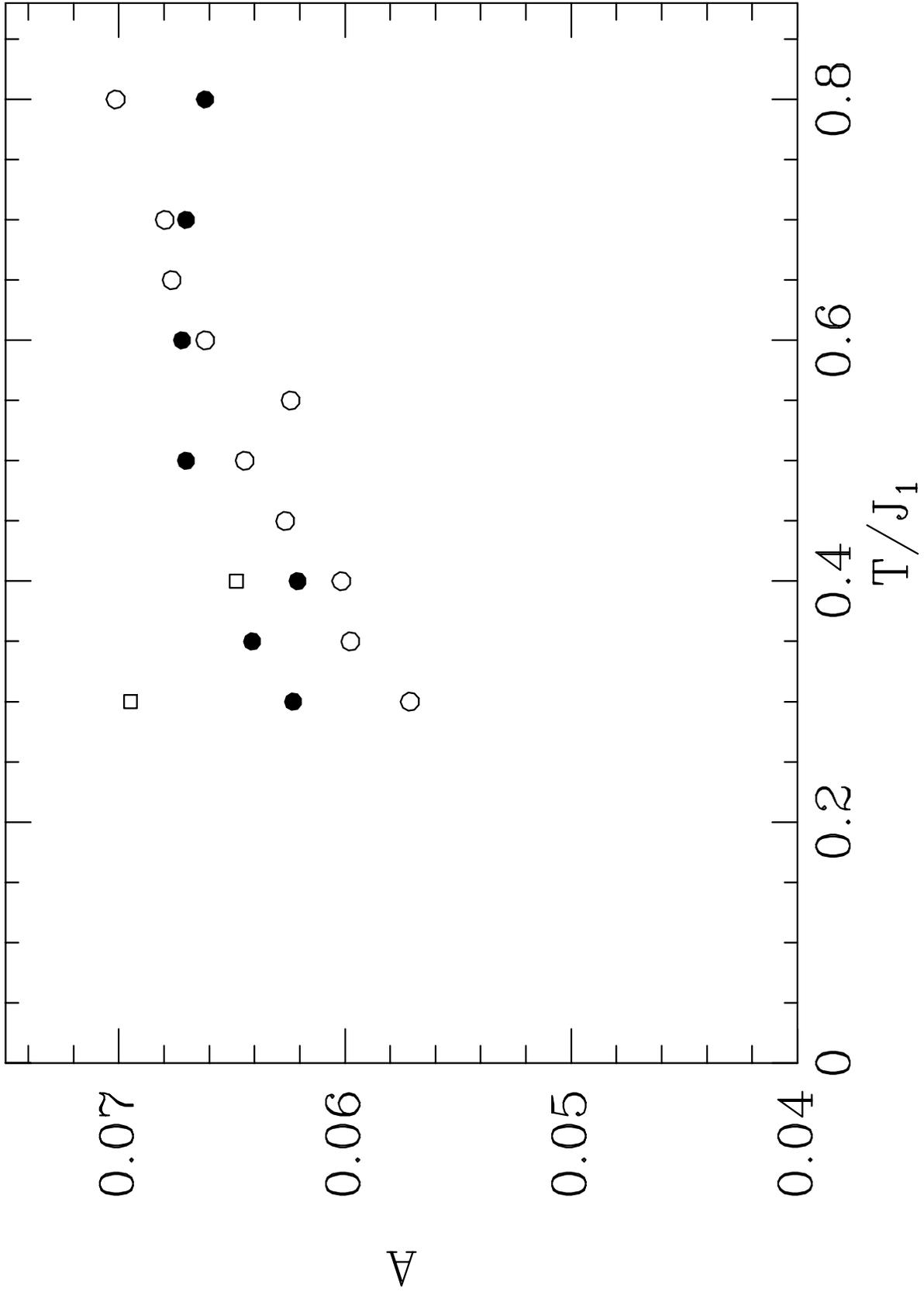
\begin{figure}
\caption{QMC results for the prefactor, $A$, in the correlation function
for interplane couplings $J_2/J_1=2.50$ (open circles), $2.55$
(solid circles), and $2.60$ (open squares).
}
\label{figa}
\end{figure}
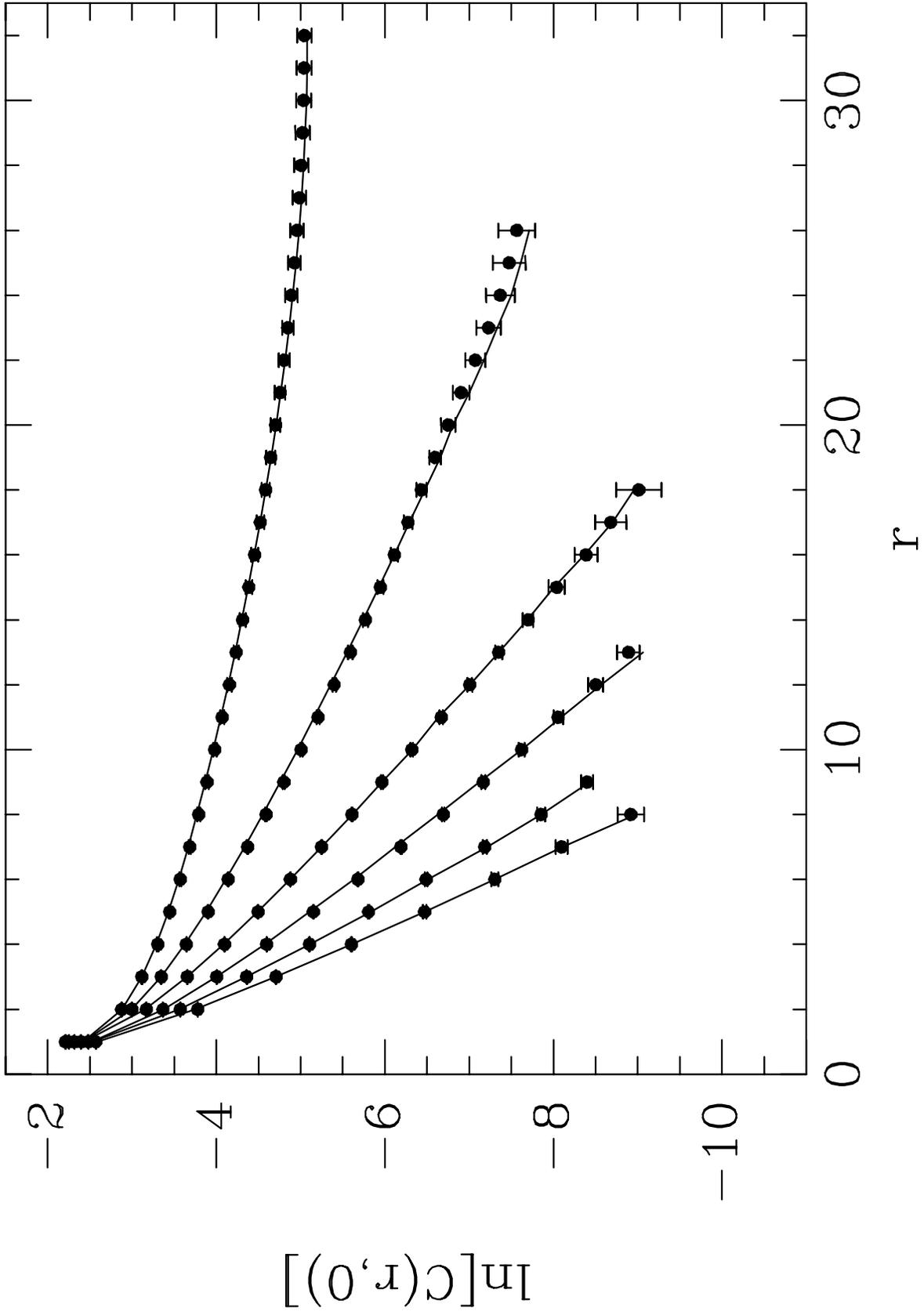
\begin{figure}
\caption{QMC results for the spin correlation function
$C(\vec r)= (-1)^{r}
\langle S^z_{r}S^z_{0}\rangle$ along $\vec r = (r,0)$
for a single-layer antiferromagnet
 at temperatures $T=0.3-0.8J_1$ (the top curve is for
$T = 0.3J_1$). The solid curves are fits to
$G(r)+G(L-r)$, with $G(r)$ given by (\protect\ref{grrc}).
}
\label{figrc}
\end{figure}


\begin{references}

\bibitem[(a)]{abo} On leave from Department of Physics, {\AA}bo Akademi,
{\AA}bo, Finland.

\bibitem{photo} J.\ M.\ Tranquada, G.\ Shirane, B.\ Keimer, S.\
Shamoto, and M.\ Sato, Phys.\ Rev.\  B {\bf 40}, 4503 (1989);
H. Ding et al, preprint.
\bibitem{Mil93} A.\ J.\ Millis and H.\ Monien, Phys.\ Rev.\ Lett.\
{\bf 70}, 2810 (1993); Phys. Rev. B {\bf 50}, 16606 (1994).
\bibitem{lee} M. Ubbens and P.A. Lee,
Phys. Rev. B {\bf 50}, 438 (1994); L.B. Ioffe et al, JETP Lett, {\bf 59}, 65
(1994); K. Kuboki and P.A. Lee, preprint.
\bibitem{CHN}  S.\ Chakravarty, B.\ I.\ Halperin, and D.\ R.\
Nelson, Phys.\ Rev.\ Lett.\ {\bf 60}, 1057 (1988); Phys.\ Rev.\ B
{\bf 39}, 2344 (1989).
\bibitem{CSY} A.\ V.\ Chubukov, S.\ Sachdev, and J.\ Ye, Phys.\
Rev.\ B {\bf 49}, 11919 (1994).
\bibitem{birg}  M. Greven et al, Phys. Rev. Lett. {\bf 72}, 1096 (1994).
\bibitem{CSS} A.\ V.\ Chubukov, S.\ Sachdev, and A.\ Sokol, Phys.\
Rev.\ B {\bf 49}, 9052 (1994).
\bibitem{Elstner} A. Sokol, R. Glenister and
R.R.P. Singh, Phys. Rev. Lett {\bf
72}, 1549 (1994); N. Elstner et al, preprint.
\bibitem{San95}  A.\ W.\ Sandvik and D.\ J.\ Scalapino, preprint.
\bibitem{berry} F.\ D.\ M.\ Haldane, Phys.\ Rev.\ Lett.\
{\bf 61}, 1029 (1988); N.\ Read and S.\ Sachdev, Phys.\ Rev.\
B {\bf 42}, 4568 (1990).
\bibitem{San94}  A.\ W.\ Sandvik and D.\ J.\ Scalapino, Phys.\ Rev.\
Lett.\ {\bf 72}, 2777 (1994)
\bibitem{dirk} A. V. Chubukov and D. Morr, in preparation.
\bibitem{qmc} A. W. Sandvik J. Phys. A{\bf 25}, 3667 (1992).
\bibitem{Hid92} K.\ Hida, J.\ Phys.\ Soc.\ Jpn.\ {\bf 61}, 1013
(1992)
\bibitem{San94_2} A.\ W.\ Sandvik and M. Vekic, preprint.
\bibitem{Sand95}  A.\ W.\ Sandvik, R.R.P. Singh and A. Sokol, unpublished.
\bibitem{tyc} S.\ Tyc, B.\ I.\ Halperin, and S.\ Chakravarty,
Phys.\ Rev.\ Lett.\ {\bf 62}, 835 (1989)
\bibitem{cr} P. Biscari, M. Campostrini and P. Rossi,  Phys. Lett B {\bf 242},
225 (1990).
\bibitem{hasen}  P. Hasenfratz and F. Niedermyer, Phys. Lett. B {\bf 268}, 231
(1991).
\bibitem{Iga92}  see e.g., J. Igarashi, Phys. Rev. B {\bf 46}, 10763 (1992).
\bibitem{DM} H.Q. Ding and M. Makivic, Phys. Rev. B {\bf 43}, 3662 (1990).
\end{references}
\end{document}